\documentclass[useAMS,usenatbib]{mn2e}
\usepackage{graphicx}

\title[Discovery of a short orbital period in the Supergiant Fast X-ray Transient IGR J16479$-$4514]{Discovery of a short orbital period in the Supergiant Fast X-ray Transient IGR J16479$-$4514}

\author[Chetana Jain, Biswajit Paul and Anjan Dutta]{Chetana Jain$^{1,2}$\thanks{E-mail: chetanajain11@gmail.com}, Biswajit Paul$^{2}$ and Anjan Dutta$^{1}$\\
$^{1}$Department of Physics and Astrophysics, University of Delhi,  Delhi 110007, India\\
$^{2}$Raman Research Institute, Sadashivnagar, C. V. Raman Avenue, Bangalore 560080, India}
\begin{document}

\pagerange{\pageref{firstpage}--\pageref{lastpage}} \pubyear{2009}
\maketitle
\label{firstpage}

\begin{abstract}
We report here discovery of a 3.32 day orbital period in the Supergiant Fast X-ray Transient (SFXT) source IGR J16479$-$4514. Using the long term light curve of this source obtained with \emph{Swift}-BAT in the energy range of 15$-$50 keV, we have clearly detected an orbital modulation including a full eclipse of duration $\sim$ 0.6 day. In the hard X-ray band of the BAT instrument, the eclipse ingress and egress are rapid. We have also used the long term light curve obtained with the \emph{RXTE}-ASM in the energy range of 1.5$-$12 keV. Taken independently, the detection of orbital modulation in the \emph{RXTE}-ASM light curve is not significant. However, considering a clear detection of orbital modulation in the BAT light curve, we have used the ASM light curve for a more precise determination of the orbital period. IGR J16479$-$4514 has the shortest orbital period among the three SFXTs with measured/known orbital period. We discuss the implication of a short orbital period with the various mechanisms proposed to explain the transient nature of this class of sources. 
\end{abstract}

\begin{keywords}
X-ray: Neutron Stars - X-ray Binaries: individual (IGR J16479$-$4514)
\end{keywords}

\section{Introduction}

High Mass X-ray binaries (HMXBs) are binary stellar systems with a compact object and an early type massive star. Some HMXBs are the brightest X-ray sources in the sky. HMXBs display a wide range of X-ray behaviour and luminosities. The X-ray emission varies from a persistent nature to a transient behaviour. Outbursts occur on timescales ranging from hours to months, and some sources also show short term flaring activity. Periodic modulation and eclipses are also seen in many HMXBs. The HMXBs can be split into two main categories as the companion star can be either a Be star or an OB supergiant (van Paradijs 1983). The majority of HMXBs ($\sim 80 \%$) belong to the first class of Be binary systems (Kaper, van der Meer $\&$ Tijani 2004). In Be system, mass transfer is effected by intermittent mass ejection from the equatorial region of a rapidly rotating Be star. The compact object is a neutron star and is typically in a wide (P$_{orb} \simeq$ 20$-$100 days) and eccentric orbit (e $\simeq$ 0.3$-$0.5) (Liu, van Paradijs $\&$ van den Heuvel 2000; Negueruela $\&$ Coe 2002). The X-ray emission is highly variable, ranging from complete absence to giant transients lasting weeks to months. In contrast to Be-star binaries, the supergiant systems are bright persistent sources in which the compact object is in orbit around an OB-supergiant (Walter et al. 2006). Accretion is powered either by the stellar wind from the supergiant and/or through Roche lobe overflow. The neutron stars orbiting wind-fed supergiant binaries are generally less luminous (L$_{x} \sim 10^{35} - 10^{37}$ ergs s$^{-1}$). The orbital periods of the supergiant systems are shorter than the Be-star systems and the orbits are more circular. However, during the past few years, the IBIS instrument (Ubertini et al. 2003) onboard the INTEGRAL $\gamma$ ray satellite (Winkler et al. 2003) has discoverd many new supergiant systems that occasionally exhibit a fast X-ray transient activity. These systems were termed as Supergiant Fast X-ray Transients (SFXTs) (Negueruela et al 2006; Smith et al. 2006).     

SFXTs are characterised by short and bright X-ray flares which last for a few hours and reach a peak luminosity of $\sim$ 10$^{36}$ ergs s$^{-1}$ (Negueruela et al. 2006; Smith et al. 2006). This new class of X-ray binaries is probably a connecting link between the Be and OB systems. They harbor a supergiant companion, similar to persistent accreting pulsars, but show bright X-ray emission only during short outbursts. These systems show recurrent outbursts, but orbital period has been measured in only two SFXTs, IGR J11215$-$5952 (Romano et al. 2007; Sidoli et al. 2007) and SAX J1818.6$-$1703 (Bird et al. 2009; Zurita Heras $\&$ Chaty 2009). Several models have been suggested to explain the origin of bright short duration flares. Bozzo, Falanga $\&$ Stella (2008a) explained that flares occur due to interaction of the inflowing wind with the neutron star magnetosphere. in't Zand (2005) proposed that the flares are due to sudden accretion of small ejections originating in a clumpy wind from the supergiant donor. However, Romano et al. (2007) showed that the accretion phase persists for a longer duration and the sources are mostly at an intermediate flaring level with an X-ray luminosity of 10$^{33} - 10^{34}$ ergs s$^{-1}$. Negueruela et al. (2006) suggested that the low rate of emission ($\sim 10^{32} - 10^{33}$ ergs s$^{-1}$) during quiescence in SFXTs is due to the fact that these systems have wider orbits as compared to persistent supergiant HMXBs, due to which the compact object accretes from a less dense environment. The quiescent state of SFXTs is characterised by a soft-spectrum and an X-ray luminosity of 10$^{32}$ ergs s$^{-1}$. SFXTs are expected to be hosting a neutron star, but pulsations have been detected only in two such systems, namely, AX J1841.0$-$0536 (4.74 s: Bamba et al. 2001) and IGR J11215$-$5952 (186.78 s: Swank, Smith $\&$ Markwardt 2007). Pulsations have also been detected in intermediate SFXTs: IGR J16465$-$4507 (227 s: Lutovinov et al. 2005; Walter et al. 2006) and IGR J18483$-$0311 (21 s: Sguera et al. 2007).

The hard X-ray transient IGR J16479$-$4514 was discovered with the INTEGRAL satellite in August 2003 (Molkov et al. 2003), with an average flux of $\sim$ 12 mCrab in the 18$-$25 keV band and $\sim$ 8 mCrab in the 25$-$50 keV band. Several fast flares were later reported from the IBIS/ISGRI observations by Sguera et al. (2005, 2006). The bright variable X-ray source is associated with a supergiant (O8.5I) star, located at a distance of $\sim$ 4.9 kpc (Chaty et al. 2008; Rahoui et al. 2008). IGR J16479$-$4514 was regularly monitored with \emph{Swift}-XRT, to study the quiescent behaviour and outburst properties (Romano et al. 2008). The X-ray emission is highly variable on timescales from seconds to weeks, both during outbursts and in quiescence (Sidoli et al. 2008). The spectrum during the bright flares is well described with an absorbed power-law model with a photon index of 0.98 (Romano et al. 2008). However, Romano et al. (2008) found that a high energy cut off powerlaw model fits well to the wide band spectrum. During the fainter emission, the spectrum was well modeled by an absorbed Comptonised spectrum, with an intrinsic column density of 7.7 $\times$10$^{22}$ cm$^{-2}$. Similar spectral features are also seen in the persistent accreting X-ray pulsars. Typical SFXTs have low quiescent X-ray luminosities of about 10$^{32}$ erg s$^{-1}$, as compared to persistent supergiant systems, which have X-ray luminosities $\sim$ 10$^{36}$ erg s$^{-1}$. In the case of IGR J16479$-$4514, Sguera et al. (2008) determined a quiescent X-ray luminosity of $\sim$10$^{34}$ erg s$^{-1}$, which is about 2 orders of magnitude higher than that of a typical SFXT. The source was observed in a low-emission phase in 2004 March with \emph{XMM-Newton} (Walter et al. 2006). The recurrent and short outbursts, the spectral properties and the association with an O-type supergiant star, confirm that IGR J16479$-$4514 is a member of the growing class of SFXTs. In addition to these features, Bozzo et al. (2008b) detected an obscuration of the X-ray source by the supergiant companion. This indicate that IGR J16479$-$4514 is possibly an eclipsing SFXT. The eclipsing nature was further supported by an increase in the EW of the iron fluorescent line at $\sim$ 6.5 keV. 

Here we report the timing analysis of the Supergiant Fast X-ray Transient source, IGR J16479$-$4514. Using the data obtained from \emph{Swift}-BAT and \emph{RXTE}-ASM observations, we have discovered a 3.32 d orbital period with clear detection of orbital modulation.  

\section{Observations and analysis}

We have analysed data from the \emph{Swift} observatory (Gehrels et al. 2004) which was launched in 2004. The scientific payload consists of a wide field instrument, the gamma ray Burst Alert Telescope (BAT; Barthelmy et al. 2005) and two co-aligned narrow field instruments: X-ray Telescope (XRT, Burrows et al. 2005) operating in the 0.2-10 keV energy band and the Ultraviolet/Optical Telescope (UVOT, Roming et al. 2005). The 15$-$50 keV \emph{Swift}-BAT light curve was obtained from the BAT Transients Monitor Program (Krimm et al. 2006) and the observations covered the time range from MJD 53413 to MJD 54857. IGR J16479$-$4514 was also monitored regularly by the All Sky Monitor (ASM) on board Rossi X-ray Timing Explorer (\emph{RXTE}). The ASM (Levine et al. 1996) comprises three wide-field scanning shadow cameras (SSCs) which are mounted on a rotating boom. The SSC's are rotated in a sequence of $``$dwells$"$ with an exposure typically of 90 s, so that most of the sky can be covered in one day. The dwell data are also averaged for each day to yield a daily-average. The data used here covered the time between MJD 50088 to MJD 54756. 

The long term light curves obtained from \emph{Swift}-BAT and \emph{RXTE}-ASM, were corrected for the earth motion using \emph{earth2sun} tool of the HEASARC software package $``$FTOOLS" ver6.5.1. We searched for the orbital period using the ftool - \emph{efsearch}, which folds the light curve with a large number of trial periods around an approximate period. The folded light curves for each trial period are fitted to a constant, and $\chi^{2}$ is determined. The trial period corresponding to the maximum $\chi^{2}$ represents the true period in the light curve, if any. For the \emph{Swift}-BAT light curve, we searched for period over a range of 0.1 to 20 days and found a peak at 286796 s and multiples of this period. The top panel of Figure 1 shows the \emph{efsearch} results from the \emph{Swift}-BAT light curve, centered at the peak $\chi^{2}$. A gaussian fit around the peak yielded a period of 286796 $\pm$ 89 s (3.3194 $\pm$ 0.0010 d) (inset of Figure 1). This is most likely to be the orbital period of the binary system and is similar to many other supergiant systems, such as, 4U 1538$-$52 (3.73 d: Becker et al. 1977), 4U 1700$-$37 (3.41 d: Jones et al. 1973), IGR J18027$-$2016 (4.6 d: Augello et al. 2003), Cen X-3 (2.09 d: Giacconi et al. 1971) and SMC X-1 (3.89 d: Schreier et al. 1972). Considering a clear detection of orbital modulation from the BAT light curve, we have used the ASM light curve for a more precise determination of the orbital period. The period search in the \emph{RXTE}-ASM light curve was done on a narrow period range and the highest peak in the $``$efsearch" result on the ASM light curve is at 286816 s (bottom panel of Figure 1) but the detection significance is poor in the energy band (1.5$-$12 keV) of the \emph{RXTE}-ASM. We also created the Lomb-Scargle periodogram by means of the fast implementation of Press $\&$ Rybicki (1989) and Scargle (1982) technique on the \emph{Swift}-BAT light curve. The long term \emph{Swift}-BAT light curve covered a span of 1335 d, with 15495 discrete pointings covering a total exposure time of $\sim$124 days. Figure 2 shows the Lomb-Scargle periodogram created from the entire \emph{Swift}-BAT light curve. A clear peak is seen in the periodogram corresponding to a frequency of 0.3012 d$^{-1}$, i.e., a period of 3.32 d, with a false alarm probability of 5e-14.  
 
\begin{figure}
\includegraphics[height=3.0in, angle=-90]{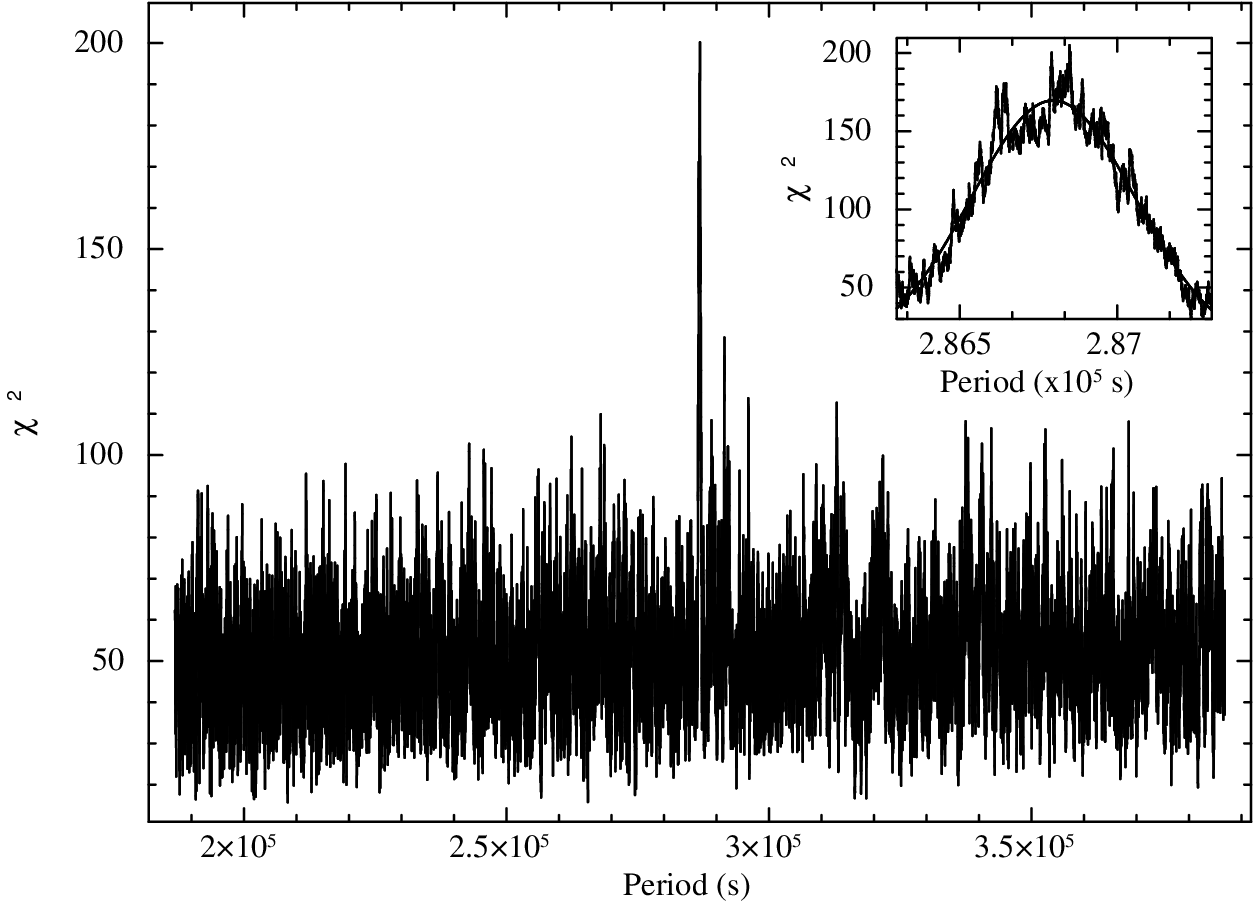}
\includegraphics[height=3.0in, angle=-90]{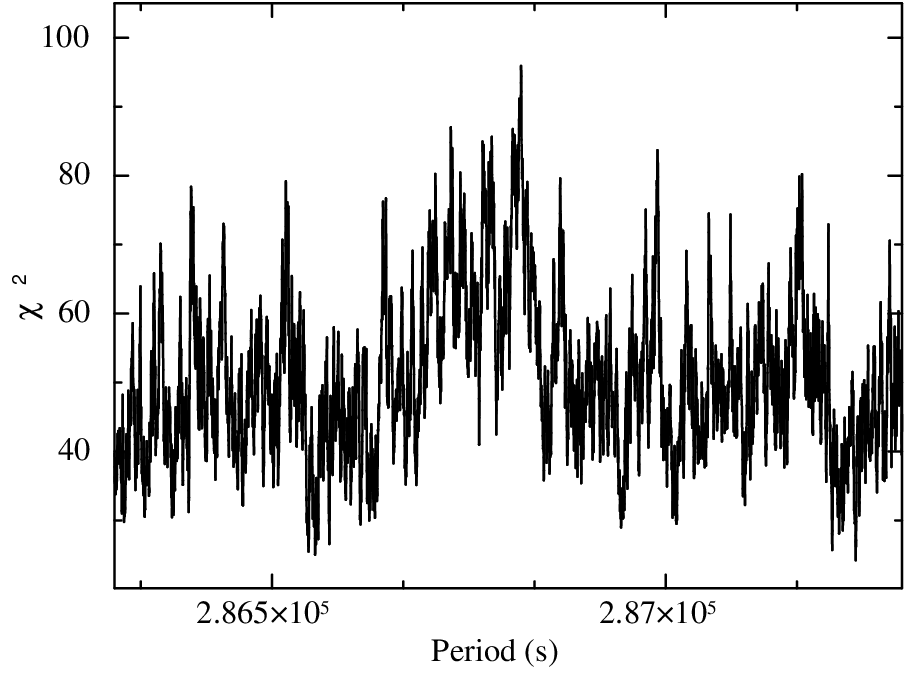}
\caption{Results from $``$efsearch" on the light curve of SFXT IGR J16479$-$4514. The top panel shows the result from the \emph{Swift}-BAT observation. The bottom panel shows the result from the \emph{RXTE}-ASM observation and the peak corresponds to a period of 286816 s. The inset figure in the top panel shows the $``$efsearch" results from \emph{Swift}-BAT light curve over the same time range as covered by the \emph{RXTE}-ASM light curve in the bottom panel. The solid line represents the best fit gaussian curve with the centre at 286796 s.}
\end{figure}

\begin{figure}
\includegraphics[height=3.0in, angle=-90]{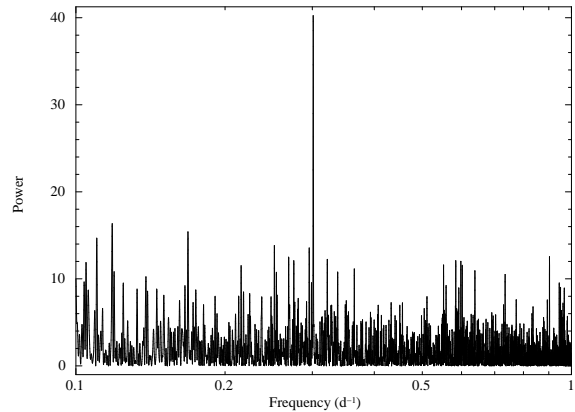}
\caption{Lomb-Scargle periodogram for the 15$-$50 keV \emph{Swift}-BAT light curve of IGR J16479$-$4514. The peak in the periodogram corresponds to a frequency of 0.3012 d$^{-1}$, i.e. 3.32 d.}
\end{figure}

Figure 3 shows the folded light curve IGR J16479$-$4514, obtained from \emph{Swift}-BAT and \emph{RXTE}-ASM observations. From the \emph{RXTE}-ASM observations, we have also folded the 5$-$12 keV light curve. The folded light curves shows clear orbital modulation. A sharp eclipse is seen from the \emph{Swift}-BAT folded light curve. From a long \emph{XMM-Newton} observation, Bozzo et al. (2008b) detected a sudden reduction in the X-ray flux and they proposed it to be due to an eclipse ingress. From the \emph{Swift}-BAT observations (which covered the time range analyzed by Bozzo et al. (2008b)), we have also detected a $\sim$ 0.6 d eclipse at the same orbital phase as reported by Bozzo et al. (2008b). The arrow in the bottom panel of Figure 3 shows the ingress phase of the eclipse as reported by Bozzo et al. (2008b). The long term average count rate, during the out-of-eclipse phase of IGR J16479$-$4514 for the \emph{Swift}-BAT light curve is 1.264 $\times$ 10$^{-3}$ counts/s (5.7 mCrab). For the entire \emph{RXTE}-ASM energy range, the out-of-eclipse count rate is 0.17 counts/s (2.2 mCrab), while for the energy range 5$-$12 keV, it is 0.07 counts/s (2.8 mCrab). The \emph{Swift}-BAT light curve, when plotted with a binsize equal to the orbital period of the system, has a total of 402 data points and 10 (6) of them are above 3 (4) $\sigma$ level. For the entire light curve, the source is detected in the \emph{Swift}-BAT (\emph{RXTE}-ASM) with a signal to noise ratio of 18 (11). The ASM orbital profile is not as significant as observed from the \emph{Swift}-BAT observations. We emphasize the presence of the eclipse at the same phase as obtained from the \emph{Swift}-BAT data and the duration of the eclipse from the two observations is also equal. The clear detection of orbital modulation shows that long term monitoring with \emph{Swift} can yield intersting results even for faint sources. Good sensitivity of \emph{Swift} combined with broad band coverage over a long timescales is crucial to determine the periodic/non-periodic phenomena in such sources. 

\begin{figure}
\includegraphics[height=3.0in, angle=-90]{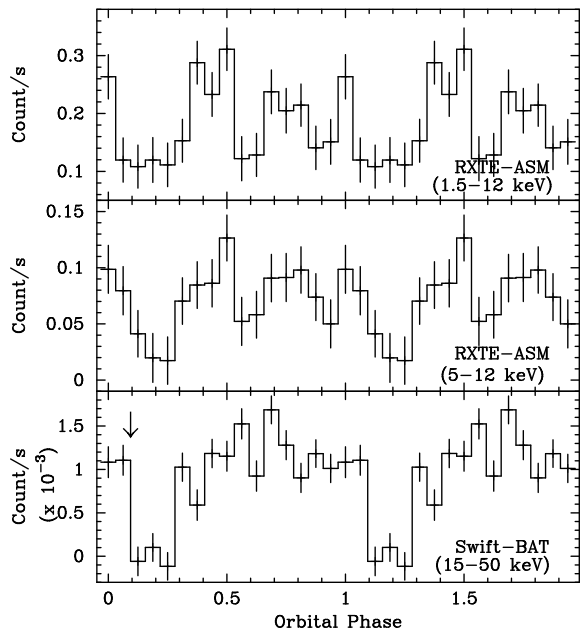}
\caption{Light curves from data obtained with \emph{Swift}-BAT and \emph{RXTE}-ASM, folded with 16 phasebins per orbit. The light curves were folded with a period of 286816 s. The top panel is the folded light curve from the entire energy range of the \emph{RXTE}-ASM data. The 5$-$12 keV folded light curve of the \emph{RXTE}-ASM observation is shown in the middle panel. The folded light curve obtained with \emph{Swift}-BAT is shown in the bottom panel. Here, the arrow corresponds to the ingress phase of the eclipse as determined by Bozzo et al. (2008b) from the \emph{XMM-Newton} observations}
\end{figure}

\section{Discussion}

Using the \emph{Swift}-BAT and \emph{RXTE}-ASM long term light curves, we have discovered an orbital X-ray modulation in the Supergiant Fast X-ray transient source, IGR J16479$-$4514. We have found an orbital period of 286796 $\pm$ 89 s (3.3194 $\pm$ 0.0010 d) and an eclipsing phase lasting $\sim$ 0.6 d.   

To date, the orbital periodicity is known only in two other SFXTs - SAX J1818.6$-$1703 (Bird et al. 2009; Zurita Heras $\&$ Chaty 2009) and IGR J11215$-$5952 (Sidoli, Paizis $\&$ Mereghetti 2006). However, the periodicity observed in these systems is considerably longer than that observed in IGR J16479$-$4514. While IGR J11215$-$5952 shows outbursts on a $\sim$ 165 d period, SAX J1818.6$-$1703 is known to exhibit a periodicity of $\sim$ 30 d. 

The frequency of occurence of long and bright flares mainly depends on the geometry of the system, wind characteristics and probably some parameter of the compact object. Different mechanisms have been proposed to explain the flaring activity in SFXTs. The long periodicity observed in IGR J11215$-$5952 is in sync with the accretion model suggested by Negueruela et al. (2006), in which they proposed that SFXTs have wide and eccentric orbits due to which the quiescent X-ray emission is very low. However, the quiescent luminosity of IGR J16479$-$4514 is two orders of magnitude higher than that of a typical SFXT (Negueruela et al. 2006; Sguera et al. 2008). This indicate that compact object remains close to the supergiant donor star and accretes a significant amount of material, both during an outburst and in quiescence. This in turn implies that the orbital period of the system should be small, otherwise quiescent phase should be longer with a lower X-ray luminosity. This is supported by our detection of an orbital period of 3.32 d. 

in't Zand (2005) had suggested the $``$clumpy winds" model to explain the occurence of short and bright flares. They had proposed that the flaring activity is related to the structure of the supergiant donor rather than to the nature of the compact object. According to their model, the radiation-driven wind from the donor star is composed of dense clumps with mass of the order of 10$^{19}$ - 10$^{20}$ g (Howk et al. 2000). The interaction of the compact object with the dense clumps formed in the massive wind leads to increased accretion rate and hard X-ray emission (Leyder et al. 2007). Oskinova et al. (2007) also proposed that optically thick macro clumps are present within the wind from the hot supergiant companions. From the hard X-ray variability of SFXT systems, Walter $\&$ Zurita Heras (2007) determined the luminosity, frequency and duration of flares in SFXTs and derived the wind clump parameters. They proposed that the orbital radius of the compact object is relatively large ($\sim$ 10 R$_{\ast}$ and the clump density varies with the orbital radius. 

Sidoli et al. (2007) proposed that the winds are anisotropic and outbursts occur due to enhanced accretion onto the neutron star when it crosses the equatorial disk component from the supergiant. However, it should be noted that in a short orbital period binary system, the accretion onto the compact object should continue for a large fraction of the orbit or the entire orbit. The average mass accretion rate should also be high and yield a relatively higher quiescent luminosity. Grebenev $\&$ Sunyaev (2007) proposed that due to propeller effect, the compact object do not emit during quiescence, but are active when the wind density increases. However, since the magnetosphere is not an ideal propeller, therefore, some fraction of the accreting matter is able to fall on the magnetic poles and quiescent emission level is detectable. In a unified $``$clumpy winds" model, Negueruela et al. (2008) proposed that different orbital configurations lead to different subclasses of supergiant systems. According to their model, in  short orbital period supergiant systems, neutron star is embedded in a quasi-continuous wind and receives a significant fraction of clumps from the companion. Whereas, in long orbital period systems, the wind fills a large volume and therefore, the clump density is small. Bozzo et al. (2008a) also explained the flaring mechanism in the context of gated accretion. They proposed that accretion onto the compact object is inhibited when the magnetospheric radius becomes larger than the corotation radius. They showed that the large luminosity swings between the quiescent and outburst phase are due to the interaction of massive infalling wind with the neutron star magnetosphere. This in turn depends on the spin period of the neutron star and the magnetic field. However, the nature of the compact object is still unknown in a majority of SFXTs. Moreover, in a short orbital period binary with an eclipse lasting for $\sim$ 20$\%$ of the period, it is intersting to explore whether gated accretion mechanism can occur at a few stellar radii where the matter density and radiation pressure are high. Further, in view of the results presented here, it is also important to explore the model put forth by Negueruela et al. (2008), specially in the case of short orbital period systems. The unified $``$clumpy'' wind model suggests that the short period systems should be bright for most of the time, but the SFXT IGR J16479$-$4514 spends most of the time in a quiescent phase, with luminosity well below the persistent bright state of normal supergiant systems.    

The observed short orbital period and the orbital modulation opens up many questions regarding the true nature of the Super Fast X-ray Transients. SFXTs share many properties with persistent accreting X-ray pulsars in supergiant HMXBs. The spectral properties of SFXTs are similar to those of accreting X-ray pulsars and the compact object is expected to be a neutron star. But, pulsations have been detected only from two such systems. In particular, the quiescent and long term average luminosity of IGR J16479$-$4514 is higher that other SFXTs but it is fainter as compared to persistent HMXBs. Therefore, a short orbital period of IGR J16479$-$4514 and a relatively higher quiescent luminosity sheds light on the probable link between the classical persistent supergiant systems (small and circular orbit) and other classical SFXTs (large and eccentric orbits). The sharp eclipse of IGR J16479$-$4514 shows that it will also be possible to determine the orbital evolution of the X-ray binary by measuring the mid-eclipse times (Jain, Paul $\&$ Dutta 2009), even if the X-ray emission does not show pulsations. Long term monitoring of all the SFXTs is therefore important to study the physical phenomena behind their unusual behaviour.

\section*{Acknowledgments}

We thank the \emph{Swift}-BAT and \emph{RXTE}-ASM teams for provision of the data. We also thank the anonymous referee for some useful comments.

\label{lastpage}
\end{document}